\documentclass[review,12pt]{elsarticle}

\usepackage{amssymb}
\usepackage{amsthm,amsmath}
\usepackage{accents}
\usepackage{graphicx}
\usepackage{epstopdf}
\newtheorem{thm}{Theorem}

\newtheorem{mydef}{Definition}
\newtheorem{asu}{Assumption}

\newtheorem{rem}{Remark}

\makeatletter
\def\widebar{\accentset{{\cc@style\underline{\mskip10mu}}}}
\makeatother
\def\proof{\noindent{\bf Proof:} }
\def\QEDclosed{\mbox{\rule[0pt]{3mm}{3mm}}}
\def\endproof{\hspace*{\fill}~\QEDclosed\par\endtrivlist\unskip}
\allowbreak \allowdisplaybreaks

\journal{Elsevier}

\begin{document}

\begin{frontmatter}

\title{Finite time anti-synchronization of complex-valued neural networks with bounded asynchronous time-varying delays\tnoteref{tt}} \tnotetext[tt]{This work was supported by the National Science Foundation of China under Grant No. 61673298, 61203149; Shanghai Rising-Star Program of China under Grant No. 17QA1404500; Natural Science Foundation of Shanghai under Grant No. 17ZR1445700; the Fundamental Research Funds for the Central Universities of Tongji University.}

\author{{Xiwei Liu\corref{lxw}}, Zihan Li}
\cortext[lxw]{Corresponding author. E-mail address: xwliu@tongji.edu.cn}

\address{Department of Computer Science and Technology, Tongji University, and with the Key Laboratory of Embedded System and Service Computing, Ministry of Education, Shanghai 201804, China}

\begin{abstract}
In this paper, we studied the finite time anti-synchronization of master-slave coupled complex-valued neural networks (CVNNs) with bounded asynchronous time-varying delays. With the decomposing technique and the generalized $\{\xi,\infty\}$-norm,
several criteria for ensuring the finite-time anti-synchronization are obtained. The whole anti-synchronization process can be divided into two parts: first, the norm of each error state component will change from initial values to $1$ in finite time, then from $1$ to $0$ in fixed time. Therefore, the whole time is finite. Finally, one typical numerical example is presented to demonstrate the correctness of our obtained results.
\end{abstract}


\end{frontmatter}


\section{Introduction}
\nocite{*}
Neural networks have been widely studied in the last thirty years and found a large number of applications tightly associated with their dynamical behaviors in many fields, such as signal processing, pattern recognition, optimization problems, and associative memories \cite{chua88}-\cite{koho88}. However, although real-valued neural networks(RVNNs) have achieved great success in many areas, they likely perform worse in some physical related applications, such as 2D affine transformation. As an extension of RVNNs, complex-valued neural networks(CVNNs) have complex-valued states, complex-valued connection weights, and complex-valued activation functions, which make them more complicated. By virtue of the characteristic of complex number, CVNNs can be applied to many physical systems related with electromagnetic, quantum waves, ultrasonic, light and so on. Moreover, CVNNs make it possible to deal with the problems which simple RVNNs cannot solve. For example, as far as we know, it is infeasible to solve the XOR problem with only one signle real-valued neuron, but it can be solved with the complex-valued one\cite{nitta2003}. Besides, it is natural to deduce CVNNs to more complicated quaternion-valued neural networks (QVNNs). Therefore, many scholars are attracted to study the dynamical behaviors and properties of CVNNs, see \cite{hu12},\cite{zhou13}-\cite{gong15b},\cite{liu15,liu18,zhou17}.

Synchronization (SYN), which is a special case of dynamical behaviors, has been extensively studied in the recent past because of its significant role in combinatorial optimization, image processing, secure communication \cite{yang97} and many other fields since it was proposed by Pecora and Carroll \cite{pecora90}. Under the concept of ``drive-response'', various kinds of SYN have been put forward so far, such as generalized SYN \cite{rulkov95}, phase SYN \cite{rosen1996}, lag SYN \cite{rosen97} and so on. In fact, there is another interesting phenomenon in chaotic oscillators, anti-synchronization (A-SYN), when A-SYN happens, the sum of two correspond state vectors can converge to zero. It can be used in many fields. For example, in communication system, the system's security and secrecy can be deeply strengthened by transforming from SYN and A-SYN periodically in the process of digital signal transmission \cite{wang18}. Hence, further study of A-SYN for dynamical systems is of high significance in both theory and practice \cite{hu2005,wu2013,liuzhu2018}.

On the other hand, in physical realization, time delay is inevitable owing to the time cost on amplifier switching and information transmission between different nodes, and it can cause undesirable impact theoretically. Thus, synchronization problem under time delay is also a hot research topic \cite{lu04}-\cite{zhang18}. In \cite{lu04,lu06}, the authors propose a new way to investigate the SYN of a class of linearly coupled ordinary differential systems. In \cite{lu11}, the global exponential SYN of linearly coupled neural networks with impulsive disturbances is solved by using differential inequality method. In \cite{wu2011}, the exponential SYN of memristor neural networks (MNN) with time-varying delays is proved based on fuzzy theory, while \cite{zhang13} solves the exponential A-SYN of MRNN with bounded delays by using differential inclusions and inequality technique. For CVNN, the SYN problem is much more difficult to solve than that of RVNN, and the biggest challenge is how to choose an appropriate activation function \cite{zhou13}. According to Liouville's theorem, any regular analytic function cannot be bounded unless it reduces to a constant. Thus, activation functions in CVNNs cannot be bounded and analytic simultaneously. One common technique is to decompose the complex-valued activation functions to its real and imaginary parts, as a result, the original CVNNs are separated into double RVNNs, many results have been obtained by applying this method, see \cite{zhou13}-\cite{liu18}. In \cite{rakki15a,gong15b}, the authors investigate the $\mu$-stability of CVNNs with unbounded time-varying delays when $f(z)$ can be expressed as $f(z)=f^R(x,y)+if^I(x,y)$, where $z=x+iy$. \cite{gong15a} considers the exponential stability problem for CVNN with time-varying delays, sufficient criteria are obtained based on the matrix measure method and Halanay inequality method, and in \cite{bao15}, the exponential SYN and A-SYN problems of complex-valued MNN with bounded delays are also solved with these two mathematical tools. \cite{liu15} investigates the exponential stability for CVNNs with asynchronous time delays by decomposing and recasting an equivalent RVNN, some sufficient conditions are given under three generalized norms. \cite{liu18} studies the A-SYN of complex-valued MNN with bounded and derivable time delays.

It should be noted that, the SYN problems presented above are under the concept of classic asymptotic SYN. In fact, based on the convergence time, synchronization can be divided into another three types: finite time SYN, fixed time SYN \cite{bhat98}. Compared with asymptotic SYN, they are more important and easier to be realized and verified in real situations. In \cite{lu16}, the authors reveal the essence of finite time and fixed time convergence by discussing the typical function $\dot{t}(V) = \mu^{-1}(V)$. In \cite{zhou17}, the authors investigate the problem of finite time SYN for CVNNs with mixed delays and uncertain perturbations. In \cite{wang16}, the finite time A-SYN of MNN with stochastic perturbations is addressed by using differential inclusions and linear matrix inequalities (LMI). In \cite{wang18,wang19}, the authors investigate the finite time A-SYN of RVNNs with bounded and unbounded time-varying delays by dividing the whole process into two procedures. In \cite{zhang19}, the finite time A-SYN problem for the master-slave neural networks with bounded time delays is considered by combining two inequalities with integral inequality skills. As far as we know, there are few works devoted to the finite time A-SYN problem for CVNNs with time delays.

Motivated by the aforementioned discussions, in this paper, we aim to solve the finite time A-SYN of CVNNs with asynchronous time-varying delays with generalized $\{\xi,\infty\}$-norm, Lyapunov functional, and inequality technique. The considered master-slave CVNNs are decomposed into their real and imaginary parts respectively. By designing a proper control law, some criteria are given for the finite time A-SYN process.

In the follwing, we give the organization structure of the rest part of this paper. In Section \ref{model}, we give the model description and decomposition, apart from this, some definitions, assumptions, and notations used later are also presented. In Section \ref{main}, we give some criteria for the finite time A-SYN for our model, and the proof . In Section \ref{nu}, one detailed numerical simulation is given to demonstrate the correctness of our obtained results. Finally, we summarize this paper and discuss about our future works in Section \ref{conclude}.

\textit{Notations} Throughout this paper, $\mathbb{R}^{m\times n}$ and $\mathbb{C}^{m\times n}$ denote any $m \times n$ dimensional real-valued and complex-valued matrices, where $\mathbb{C}$ is
the set of complex numbers. For any vector $\alpha=(\alpha_j)_{1\times n}$, $j=1,\cdots,n$, denote $\alpha^T$ as the transposition of $\alpha$, and denote $|\alpha|=(|\alpha_j|)_{1\times n}$.

\section{Model description}\label{model}
At first, we present some matrices to show the property of the dot product operation between any two complex-valued numbers $a$ and $b$, where $a=a^R+ia^I,b=b^R+ib^I$.

Define a 2-dimensional matrix
\begin{align}
M&=\left(\begin{array}{cc}1&i\\i&-1\end{array}\right)=M^R+iM^I,
\end{align}
where
\begin{align*}
M^R=\left(\begin{array}{cc}1&0\\0&-1\end{array}\right),~~~M^I=\left(\begin{array}{cc}0&1\\1&0\end{array}\right).
\end{align*}
\begin{mydef}
	For any two complex numbers $a$ and $b$, denote
	\begin{align}\label{notation}
	\widehat{a} = (a^R, a^I)^T,~~~\widehat{b} = (b^R, b^I)^T,
	\end{align}
	then
	\begin{align}\label{add1}
	a\cdot b=\widehat{a}^TM\widehat{b}=\widehat{a}^TM^R\widehat{b}+i\widehat{a}^TM^I\widehat{b}
	\end{align}
	i.e.,
	\begin{align}\label{add2}
	\widehat{ab}=(\widehat{a}^TM^R\widehat{b}, \widehat{a}^TM^I\widehat{b})^T=(a^Rb^R-a^Ib^I,a^Rb^I+a^Ib^R)^T
	\end{align}
\end{mydef}
In the following paper, this property is utilized to reduce the redundancy of the calculation and representation.

Consider the following CVNN with asynchronous time-varying delays:
\begin{align}\label{master}
\dot{x}_j(t)=-d_jx_j(t)+\sum_{k=1}^na_{jk}f_k(x_k(t))
+\sum_{k=1}^nb_{jk}g_k(x_k(t-\tau_{jk}(t)))+H_j
\end{align}
where $x_j=x_j^R+ix_j^I\in \mathbb{C}$ is the state of $j$-th neuron, $j=1,\cdots,n$; $D=\mathrm{diag}(d_1, \cdots, d_n)\in\mathbb{R}^{n\times n}$ with $d_j>0$ is the feedback self-connection weight matrix; $f_j(\cdot): \mathbb{C}\rightarrow \mathbb{C}$ and $g_j(\cdot): \mathbb{C}\rightarrow \mathbb{C}$ are complex-valued activation functions without and with time delays; matrices $A=(a_{jk})\in \mathbb{C}^{n\times n}$ and $B=(b_{jk})\in \mathbb{C}^{n\times n}$ represent the complex-valued connection weight matrices without and with time delays; $\tau_{jk}(t)$ is bounded, asynchronous, and time-varying with $0\le\tau_{jk}(t)\le\tau$; $H_j\in \mathbb{C}$ is the $j$-th external input. The initial functions of (\ref{master}) are given by
\begin{align*}
&x_j(\theta)=\Phi_j(\theta)=\Phi_j^R(\theta) + i\Phi_j^I(\theta),~~~\mathrm{for}~\theta\in[-\tau,0],~~~j=1,\cdots,n
\end{align*}

For convenience, we denote
\begin{align}
&f_k^{\ell}(x; t)=f_k^\ell(x_k(t)),~g_{k\tau_{jk}}^{\ell}(x; t)=g_k^\ell(x_k(t-\tau_{jk}(t))),
\end{align}
then according to rule (\ref{add1}), CVNN (\ref{master}) can be decomposed into two equivalent RVNNs:
\begin{align}
\dot{x}_j^R&(t)=-d_jx_j^R(t)+\sum_{k=1}^n\widehat a_{jk}^TM^R\widehat f_k(x;t)+\sum_{k=1}^n\widehat b_{jk}^TM^R\widehat g_{k\tau_{jk}}(x;t)+H_j^R,\label{masterR}\\
\dot{x}_j^I&(t)=-d_jx_j^I(t)+\sum_{k=1}^n\widehat a_{jk}^TM^I\widehat f_k(x;t)+\sum_{k=1}^n\widehat b_{jk}^TM^I\widehat g_{k\tau_{jk}}(x;t)+H_j^I.\label{masterI}
\end{align}

Let (\ref{master}) be the master system, then the slave system is defined as
\begin{align}\label{slave}
\dot{y}_j(t)&=-d_jy_j(t)+\sum_{k=1}^na_{jk}f_k(y_k(t))
+\sum_{k=1}^nb_{jk}g_k(y_k(t-\tau_{jk}(t)))+H_j+u_j(t)
\end{align}
with the initial state
\begin{align*}
&y_j(\theta)=\Psi_j(\theta)=\Psi_j^R(\theta) + i\Psi_j^I(\theta),~~~\mathrm{for}~\theta\in[-\tau,0],~~~j=1,\cdots,n
\end{align*}

The control scheme $u_j(t) \in \mathbb{C}$, $j=1,\cdots,n$ is designed to be only depending on the system state at the present time and will be defined later.

Similarly, CVNN (\ref{slave}) can be decomposed into two equivalent RVNNs:
\begin{align}
\dot{y}_j^R(t)=-d_jy_j^R(t)+\sum_{k=1}^n\widehat a_{jk}^TM^R\widehat f_k(y;t)+\sum_{k=1}^n\widehat b_{jk}^TM^R\widehat g_{k\tau_{jk}}(y;t)+H_j^R+u_j^R(t),\label{slaveR}\\
\dot{y}_j^I(t)=-d_jy_j^I(t)+\sum_{k=1}^n\widehat a_{jk}^TM^I\widehat f_k(y;t)+\sum_{k=1}^n\widehat b_{jk}^TM^I\widehat g_{k\tau_{jk}}(y;t)+H_j^I+u_j^I(t).~~~\label{slaveI}
\end{align}

Denote $e_j(t)=x_j(t)+y_j(t)$ be the $j$-th component of A-SYN error between networks (\ref{master}) and (\ref{slave}), then one can get
\begin{align}\label{error}
\dot{e}_j(t)=&-d_je_j(t)+\sum_{k=1}^na_{jk}\bigg(f_k(x;t)+f_k(y;t)\bigg)\nonumber\\
&+\sum_{k=1}^nb_{jk}\bigg(g_{k\tau_{jk}}(x;t)+g_{k\tau_{jk}}(y;t))\bigg)+2H_j+u_j(t), ~~j=1,\cdots,n
\end{align}

Similarly, denote $e_j^R(t)=x_j^R(t)+y_j^R(t)$ and $e_j^I(t)=x_j^I(t)+y_j^I(t)$, then system (\ref{error}) can also be decomposed as
\begin{align}
\dot{e}_j^R(t)=&-d_je_j^R(t)+\sum_{k=1}^n\widehat a_{jk}^TM^R\bigg(\widehat f_k(x;t)+\widehat f_k(y;t)\bigg)\nonumber\\
&+\sum_{k=1}^n\widehat b_{jk}^TM^R\bigg(\widehat g_{k\tau_{jk}}(x;t)+\widehat g_{k\tau_{jk}}(y;t)\bigg)+2H_j^R+u_j^R(t),\label{errorR}\\
\dot{e}_j^I(t)=&-d_je_j^I(t)+\sum_{k=1}^n\widehat a_{jk}^TM^I\bigg(\widehat f_k(x;t)+\widehat f_k(y;t)\bigg)\nonumber\\
&+\sum_{k=1}^n\widehat b_{jk}^TM^I\bigg(\widehat g_{k\tau_{jk}}(x;t)+\widehat g_{k\tau_{jk}}(y;t)\bigg)+2H_j^I+u_j^I(t).\label{errorI}
\end{align}

As for the measurement of the A-SYN error, we choose the following generalized $\{\xi,\infty\}$-norm in this paper.
\begin{mydef}(\cite{chen2001})
For any vector $v(t)=(v_1(t),v_2(t),\cdots,v_n(t))^T\in\mathbb R^{n\times 1}$, its $\{\xi,\infty\}$-norm is defined as:
\begin{align}\label{norm}
\|v(t)\|_{\{\xi,\infty\}}=\max\{|\xi_j^{-1} v_j(t)|\},
\end{align}
where $\xi=(\xi_1,\cdots,\xi_n)^T$ with $\xi_j>0$, $j=1,\cdots,n$. Obviously, when $\xi=(1,\cdots,1)^T$, this $\{\xi,\infty\}$-norm is the conventional $\infty$-norm.
\end{mydef}

Now, we give some assumptions on the activation functions.
\begin{asu}\label{oddf}
Assume $f_k(x_k)$ and $g_k(x_k)$can be decomposed into real and imaginary part as $f_k(x_k)=f_k^R(x_k^R,x_k^I)+if_k^I(x_k^R,x_k^I)$ and $g_k(x_k)=g_k^R(x^R_k,x^I_k)+ig_k^I(x_k^R,x_k^I)$, $f_k^{\ell}$ and $g_k^{\ell}$ are all {\bf odd} functions, for $\ell=R,I$, i.e.,
\begin{align}\label{odd}
f_k^{\ell}(x_k^R,x_k^I) =-f_k^{\ell}(-x_k^R,-x_k^I),~~~~g_k^{\ell}(x_k^R,x_k^I)=-g_k^{\ell}(-x_k^R,-x_k^I).
\end{align}
\end{asu}

\begin{asu}\label{partial}	
Suppose that $f_k^{\ell}$ and $g_k^{\ell}$ are Lipschitz-continuous with respect to (w.r.t.) each component, i.e., there exist positive constants $\lambda_k^{\ell_1\ell_2}, \gamma_k^{\ell_1\ell_2}$, $\ell_1, \ell_2=R, I,$ such that
\begin{align}\label{add3}
0\le \frac{\partial{f_k^{\ell_1}}(x_k^R,x_k^I)}{\partial{x_k^{\ell_2}}}\le \lambda_k^{\ell_1\ell_2},~~~~
\bigg|\frac{\partial{g_k^{\ell_1}}(x_k^R,x_k^I)}{\partial{x_k^{\ell_2}}}\bigg|\le \gamma_k^{\ell_1\ell_2}
\end{align}
\end{asu}

Using these constants $\lambda_k^{\ell_1\ell_2}, \kappa_k^{\ell_1\ell_2}$, we can define four matrices which will be used in the following analysis:
\begin{align}\label{matrix}
\widebar\Lambda_k = \left(\begin{array}{cc}
\lambda_k^{RR}&\lambda_k^{RI}\\\lambda_k^{IR}&\lambda_k^{II}
\end{array}\right),
\widebar \Gamma_k = \left(\begin{array}{cc}
\gamma_k^{RR}&\gamma_k^{RI}\\ \gamma_k^{IR}&\gamma_k^{II}
\end{array}\right)\nonumber\\
\widetilde\Lambda_k = \left(\begin{array}{cc}
\lambda_k^{IR}&\lambda_k^{II}\\\lambda_k^{RR}&\lambda_k^{RI}
\end{array}\right),
\widetilde \Gamma_k = \left(\begin{array}{cc}
\gamma_k^{IR}&\gamma_k^{II}\\ \gamma_k^{RR}&\gamma_k^{RI}
\end{array}\right)
\end{align}

\section{Main results}\label{main}
In this section, we prove that the error systems (\ref{errorR}) and (\ref{errorI}) can achieve A-SYN in finite time.

At first, let us define the external controller $u_j(t)=u_j^R(t)+iu_j^I(t)$ as
\begin{align}
u_j^R(t) &= -\mathrm{sign}(e_j^R(t))\big[\widebar{\mu}_j|e_j^R(t)|+\widebar{\rho}_j|e_j^R(t)|^\beta+\widebar{\eta}_j\big]\label{add4}\\
u_j^I(t) &= -\mathrm{sign}(e_j^I(t))\big[\widetilde{\mu}_j|e_j^I(t)|+\widetilde{\rho}_j|e_j^I(t)|^\beta+\widetilde{\eta}_j\big]\label{add5}
\end{align}
where $0 <\beta<1$, $\widebar{\mu}_j$, $\widebar{\rho}_j$, $\widebar{\eta}_j$, $\widetilde{\mu}_j$, $\widetilde{\rho}_j$, $\widetilde{\eta}_j$ will be defined in the next theorem.

\begin{thm}\label{theorem1}
Assume Assumptions \ref{oddf} and \ref{partial} hold, error systems (\ref{errorR}) and (\ref{errorI}) will achieve A-SYN in finite time if there exists a vector
\begin{align}\label{weight}
\xi=(\xi_1,\xi_2,\cdots,\xi_n,\phi_1,\phi_2,\cdots,\phi_n)^T>0
\end{align}
such that for any $j=1,2,\cdots,n$, the following inequalities hold:
\begin{align}
\widebar \mu_{j}>&-d_j+(\{a_{{j}{j}}^R\}^{+}, \{-a_{{j}{j}}^I\}^{+})(\lambda_{{j}}^{RR},\lambda_{{j}}^{IR})^T
+\xi_{j}^{-1}\phi_{j}(|a_{{j}{j}}^R|, |a_{{j}{j}}^I|)(\lambda_{{j}}^{RI},\lambda_{{j}}^{II})^T\nonumber\\
&+\xi_{j}^{-1}\sum_{k\ne {j}}|\widehat a^T_{{j}k}|\widebar\Lambda_k(\xi_k,{\phi_k})^T+\xi_{j}^{-1}\sum_{k=1}^n|\widehat b^T_{{j}k}|\widebar \Gamma_k (\xi_k,{\phi_k})^T,\label{mubar}\\
\widetilde{\mu}_j>&-d_j+\phi_j^{-1}\xi_j(|a_{jj}^I|,|a_{jj}^R|)(\lambda_{{j}}^{RR},\lambda_{{j}}^{IR})^T+(\{a_{jj}^I\}^{+},\{a_{jj}^R\}^{+})(\lambda_{{j}}^{RI},\lambda_{{j}}^{II})^T\nonumber\\
&+\phi_j^{-1}\sum_{k\ne j}|\widehat a_{jk}^T|\widetilde\Lambda_k (\xi_k,\phi_k)^T+\phi_j^{-1}\sum_{k=1}^n|\widehat b_{jk}^T|\widetilde \Gamma_k (\xi_k,\phi_k)^T,\label{mutilde}\\
\widebar{\rho}_j>&\bigg(-d_{j}+(\{a_{jj}^R\}^{+}, \{-a_{jj}^I\}^{+})(\lambda_{j}^{RR},\lambda_{j}^{IR})^T\nonumber\\
&+(\phi_j^{-1}\xi_j)^{\frac{1}{\beta-1}}(|a_{jj}^R|, |a_{jj}^I|)(\lambda_j^{RI},\lambda_{j}^{II})^T\nonumber\\
&+\xi_{j}^{\frac{1}{\beta-1}}\sum_{k\ne j}|\widehat a^T_{jk}|\widebar\Lambda_k(\xi_k^{\frac{1}{1-\beta}},\phi_k^{\frac{1}{1-\beta}})^T-\widebar \mu_{j}\bigg)^{+},\label{rhobar}\\
\widetilde{\rho}_j>&\bigg(-d_{j}+(\{a_{jj}^I\}^{+},\{a_{jj}^R\}^{+})(\lambda_{{j}}^{RI},\lambda_{{j}}^{II})^T\nonumber\\
&+(\xi_j^{-1}\phi_j)^{\frac{1}{\beta-1}}(|a_{jj}^I|,|a_{jj}^R|)(\lambda_{{j}}^{RR},\lambda_{{j}}^{IR})^T\nonumber\\
&+\phi_{j}^{\frac{1}{\beta-1}}\sum_{k\ne j}|\widehat a^T_{jk}|\widetilde\Lambda_k(\xi_k^{\frac{1}{1-\beta}},\phi_k^{\frac{1}{1-\beta}})^T-\widetilde \mu_{j}\bigg)^{+},\label{rhotilde}\\
\widebar{\eta}_j \ge& \sum_{k=1}^n|\widehat b_{jk}^T|\widebar \Gamma_k (1,1)^T+2|H_j^R|,\label{eta}\\
\widetilde{\eta}_j \ge& \sum_{k=1}^n|\widehat b_{jk}^T|\widetilde \Gamma_k (1,1)^T+2|H_j^I|,\label{etatilde}
\end{align}
where $a^{+}=\max(0,a)$. There are two important time points, where $T_1$ denotes the first time the $\{\xi,\infty\}$-norm of the errors in (\ref{errorR}) and (\ref{errorI}) have all crossed over $1$, $T_2$ denotes the first time the error values all become $0$. Exact values of $T_1$ and $T_2$ will be given in the proof.
%
\end{thm}

\proof
	For real-valued systems (\ref{errorR}) and (\ref{errorI}), if $\sup \limits_{-\tau\le s\le 0}\big(\max\limits_{j=1,\cdots,n}|e_j^R(s)|\big) \le 1$ and $\sup \limits_{-\tau\le s\le 0}\big(\max\limits_{j=1,\cdots,n}|e_j^I(s)|\big) \le 1$, then we can deduce that $T_1=0$. Otherwise, by conditions (\ref{mubar}) and (\ref{mutilde}), we can choose a constant $\epsilon>0$ small enough so that
\begin{align}
&(\epsilon-d_{j}-\widebar \mu_{j})+(\{a_{{j}{j}}^R\}^{+}, \{-a_{{j}{j}}^I\}^{+})(\lambda_{{j}}^{RR},\lambda_{{j}}^{IR})^T
+\xi_{j}^{-1}\phi_{j}(|a_{{j}{j}}^R|, |a_{{j}{j}}^I|)(\lambda_{{j}}^{RI},\lambda_{{j}}^{II})^T\nonumber\\
&+\xi_{j}^{-1}\sum_{k\ne {j}}|\widehat a^T_{{j}k}|\widebar\Lambda_k(\xi_k,{\phi_k})^T+\xi_{j}^{-1}e^{\epsilon \tau}\sum_{k=1}^n|\widehat b^T_{{j}k}|\widebar \Gamma_k (\xi_k,{\phi_k})^T<0,\label{first1}\\
&(\epsilon-d_j-\widetilde\mu_j)+\phi_j^{-1}\xi_j(|a_{jj}^I|,|a_{jj}^R|)(\lambda_{{j}}^{RR},\lambda_{{j}}^{IR})^T+(\{a_{jj}^I\}^{+},\{a_{jj}^R\}^{+})(\lambda_{{j}}^{RI},\lambda_{{j}}^{II})^T\nonumber\\
&+\phi_j^{-1}\sum_{k\ne j}|\widehat a_{jk}^T|\widetilde\Lambda_k (\xi_k,\phi_k)^T+\phi_j^{-1}e^{\epsilon\tau}\sum_{k=1}^n|\widehat b_{jk}^T|\widetilde \Gamma_k (\xi_k,\phi_k)^T<0.\label{first2}
	\end{align}
	For all $t\ge0$, denote
	\begin{align*}
	E_1(t) =(e_1^R, e_2^R, \cdots, e_n^R, e_1^I, e_2^I, \cdots, e_n^I)^T
	\end{align*}
	with
	\begin{align}\label{xtinfty}
	\|E_1(t)\|_{\{\xi,\infty\}} = \max\Big\{\max \limits_{j=1,\cdots,n}\big\{|\xi_j^{-1}e_j^R(t)|\big\},\max \limits_{j=1,\cdots,n}\big\{|\phi_j^{-1}e_j^I(t)|\big\}\Big\},
	\end{align}
	and
	\begin{align}\label{mx(t)}
	M(E_1(t)) = \sup \limits_{t-\tau\le s\le t}\bigg(e^{\epsilon s} \|E_1(s)\|_{\{\xi,\infty\}}\bigg).
	\end{align}

Obviously, $e^{\epsilon t}|\xi_j^{-1} e_j^R(t)|\le M(E_1(t))$, and $e^{\epsilon t}|\phi_j^{-1} e_j^I(t)|\le M(E_1(t))$.
	
In the following, we only discuss the condition $e^{\epsilon t}|\xi_j^{-1} e_j^R(t)|\le M(E_1(t))$. The other case can also be discussed with the same process.
	
$(I)$ If $e^{\epsilon t}|\xi_j^{-1} e_j^R(t)|< M(E_1(t))$ for all $j=1,\cdots,n$, we know that there must be a constant $\delta_1>0$ with which $e^{\epsilon s}|\xi_j^{-1} e_j^R(s)|< M(E_1(t))$ and $M(E_1(s)\big) \le M(E_1(t))$ for $s\in(t,t+\delta_1)$.
	
$(II)$ If there exist an index $\widebar j_0$ and a time point $\widebar t_0 \ge 0$ such that
\begin{align}
e^{\epsilon \widebar t_0}|\xi_{\widebar j_0}^{-1} e_{\widebar j_0}^R(\widebar t_0)|= M(E_1(\widebar t_0)),
\end{align}
then one gets
\begin{align*}
	&\xi_{\widebar j_0}\frac{dM(E_1(t))}{dt}\bigg|_{t=\widebar t_0}
	=\frac{de^{\epsilon t}|e_{\widebar j_0}^R(t)|}{dt}\bigg|_{t=\widebar t_0}\\
	=&\epsilon e^{\epsilon \widebar t_0}|e_{\widebar j_0}^R|+e^{\epsilon \widebar t_0}\mathrm{sign}(e_{\widebar j_0}^R)\cdot \bigg\{-d_{\widebar j_0}e_{\widebar j_0}^R+\sum_{k=1}^n\widehat a_{{\widebar j_0}k}^TM^R\bigg(\widehat f_k(x;\widebar t_0)+\widehat f_k(y;\widebar t_0)\bigg)\\
	&+\sum_{k=1}^n\widehat b_{{\widebar j_0}k}^TM^R\bigg(\widehat g_{k\tau_{\widebar j_0k}}(x;\widebar t_0)+\widehat g_{k\tau_{\widebar j_0k}}(y;\widebar t_0)\bigg)+2H_{\widebar j_0}^R\\
&-\mathrm{sign}(e_{\widebar j_0}^R)\bigg(\widebar{\mu}_{\widebar j_0}|e_{\widebar j_0}^R|+\widebar{\rho}_{\widebar j_0}|e_{\widebar j_0}^R|^\beta+\widebar{\eta}_{\widebar j_0}\bigg)\bigg\}\\
\le&e^{\epsilon \widebar t_0}\bigg\{(\epsilon-d_{\widebar j_0})|e_{\widebar j_0}^R(\widebar t_0)|+\mathrm{sign}(e_{\widebar j_0}^R)\widehat a_{{\widebar j_0}{\widebar j_0}}^TM^R\bigg(\widehat f_{\widebar j_0}(x;\widebar t_0)+\widehat f_{\widebar j_0}(y;\widebar t_0)\bigg)\\
&+\sum_{k\ne {\widebar j_0}}|\widehat a_{{\widebar j_0}k}|^T|\widehat f_k(x;\widebar t_0)+\widehat f_k(y;\widebar t_0)|\\	
&+\sum_{k=1}^n|\widehat b_{{\widebar j_0}k}|^T|\widehat g_{k\tau_{\widebar j_0k}}(x;\widebar t_0)+\widehat g_{k\tau_{\widebar j_0k}}(y;\widebar t_0)|+2|H_{\widebar j_0}^R|\\
	&-\widebar \mu_{\widebar j_0}|e_{\widebar j_0}^R(\widebar t_0)| - \widebar \rho_{\widebar j_0}{|e_{\widebar j_0}^R(\widebar t_0)|}^\beta-\widebar\eta_{\widebar j_0}\bigg\}\nonumber\\
\le&e^{\epsilon \widebar t_0}\bigg\{(\epsilon-d_{\widebar j_0})|e_{\widebar j_0}^R(\widebar t_0)|+(\{a_{{\widebar j_0}{\widebar j_0}}^R\}^{+}, \{-a_{{\widebar j_0}{\widebar j_0}}^I\}^{+})(\lambda_{{\widebar j_0}}^{RR},\lambda_{{\widebar j_0}}^{IR})^T|e_{\widebar j_0}^R(\widebar t_0)|\\
&+(|a_{{\widebar j_0}{\widebar j_0}}^R|, |a_{{\widebar j_0}{\widebar j_0}}^I|)(\lambda_{{\widebar j_0}}^{RI},\lambda_{{\widebar j_0}}^{II})^T|e_{\widebar j_0}^I(\widebar t_0)|+\sum_{k\ne {\widebar j_0}}|\widehat a^T_{{\widebar j_0}k}|\widebar\Lambda_k|\widehat e_k(\widebar t_0)|\\	
&+\sum_{k=1}^n|\widehat b^T_{{\widebar j_0}k}|\widebar \Gamma_k |\widehat e_{k\tau_{\widebar j_0k}}(\widebar t_0)|-\widebar \mu_{\widebar j_0}|e_{\widebar j_0}^R(\widebar t_0)| \bigg\}\\
=&\xi_{\widebar j_0}(\epsilon-d_{\widebar j_0})e^{\epsilon \widebar t_0}\xi_{\widebar j_0}^{-1}|e_{\widebar j_0}^R(\widebar t_0)|\\
&+\xi_{\widebar j_0}(\{a_{{\widebar j_0}{\widebar j_0}}^R\}^{+}, \{-a_{{\widebar j_0}{\widebar j_0}}^I\}^{+})(\lambda_{{\widebar j_0}}^{RR},\lambda_{{\widebar j_0}}^{IR})^Te^{\epsilon \widebar t_0}\xi_{\widebar j_0}^{-1}|e_{\widebar j_0}^R(\widebar t_0)|\\
&+\phi_{\widebar j_0}(|a_{{\widebar j_0}{\widebar j_0}}^R|, |a_{{\widebar j_0}{\widebar j_0}}^I|)(\lambda_{{\widebar j_0}}^{RI},\lambda_{{\widebar j_0}}^{II})^Te^{\epsilon \widebar t_0}\phi_{\widebar j_0}^{-1}|e_{\widebar j_0}^I(\widebar t_0)|\\
&+\sum_{k\ne {\widebar j_0}}|\widehat a^T_{{\widebar j_0}k}|\widebar\Lambda_k\mathrm{diag}(\xi_k,{\phi_k})e^{\epsilon \widebar t_0}(\xi_k^{-1}|\widehat e_k^R(\widebar t_0)|,\phi_k^{-1}|\widehat e_k^I(\widebar t_0)|)^T\\
&+e^{\epsilon \tau_{\widebar j_0k}(\widebar t_0)}\sum_{k=1}^n|\widehat b^T_{{\widebar j_0}k}|\widebar \Gamma_k \mathrm{diag}(\xi_k,{\phi_k})e^{\epsilon (\widebar t_0-\tau_{\widebar j_0k}(\widebar t_0))}\cdot\\
&~~~~(\xi_k^{-1}|\widehat e_{k\tau_{\widebar j_0k}}^R(\widebar t_0)|,\phi_k^{-1}|\widehat e_{k\tau_{\widebar j_0k}}^I(\widebar t_0)|)^T-\xi_{\widebar j_0}\widebar \mu_{\widebar j_0}e^{\epsilon \widebar t_0}\xi_{\widebar j_0}^{-1}|e_{\widebar j_0}^R(\widebar t_0)|\\
\le&\xi_{\widebar j_0}\bigg\{(\epsilon-d_{\widebar j_0}-\widebar \mu_{\widebar j_0})+(\{a_{{\widebar j_0}{\widebar j_0}}^R\}^{+}, \{-a_{{\widebar j_0}{\widebar j_0}}^I\}^{+})(\lambda_{{\widebar j_0}}^{RR},\lambda_{{\widebar j_0}}^{IR})^T\\
&+\xi_{\widebar j_0}^{-1}\phi_{\widebar j_0}(|a_{{\widebar j_0}{\widebar j_0}}^R|, |a_{{\widebar j_0}{\widebar j_0}}^I|)(\lambda_{{\widebar j_0}}^{RI},\lambda_{{\widebar j_0}}^{II})^T+\xi_{\widebar j_0}^{-1}\sum_{k\ne {\widebar j_0}}|\widehat a^T_{{\widebar j_0}k}|\widebar\Lambda_k(\xi_k,{\phi_k})^T\\
&+\xi_{\widebar j_0}^{-1}e^{\epsilon \tau}\sum_{k=1}^n|\widehat b^T_{{\widebar j_0}k}|\widebar \Gamma_k (\xi_k,{\phi_k})^T\bigg\}\|e^{\epsilon \widebar t_0}E_1(\widebar t_0)\|_{\{\xi,+\infty\}}\\
\le& 0
	\end{align*}

Otherwise, $e^{\epsilon t}|\phi_j^{-1} e_j^I(t)|\le M(E_1(t))$, there also exists two cases, and the derivation procedure is similar to the content above, so it is omitted here.
	
Therefore, for all $t\ge0$, $M(E_1(t))$ is non-increasing and $M(E_1(t)) \le M(E_1(0))$, which means that
	\begin{align*}
	\min\limits_{j=1,\cdots,n}\{\xi_j^{-1}\}e^{\epsilon(t-\tau)}\sup \limits_{t-\tau\le s\le t}\big(\max\limits_{j=1,\cdots,n}|e_j^R(s)|\big)
	\le M(E_1(t)) \le M(E_1(0))
	\end{align*}
	i.e.,
	\begin{align*}
	\sup \limits_{t-\tau\le s\le t}\big(\max\limits_{j=1,\cdots,n}|e_j^R(s)|\big)
	\le \frac{\max\limits_{j=1,\cdots,n}\{\xi_j\}M(E_1(0))}{e^{\epsilon(t-\tau)}}
	\end{align*}

Thus, as time $t$ increases, $\sup \limits_{t-\tau\le s\le t}\big(\max\limits_{j=1,\cdots,n}|e_j^R(s)|\big)$ would be less than $1$. We denote $T_1^R$ as the first time point such that $\frac{\max\limits_{j=1,\cdots,n}\{\xi_j\}M(E_1(0))}{e^{\epsilon(T_1^R-\tau)}} = 1$, and $\sup \limits_{t-\tau\le s\le t}\big(\max\limits_{j=1,\cdots,n}|e_j^R(s)|\big) \le 1$ for $t \ge T_1^R$, here
	\begin{align}
	T_1^R = {\epsilon}^{-1}\ln{\bigg(\max\limits_{j=1,\cdots,n}\{\xi_j\}M(E_1(0))\bigg)} + \tau. \label{t1r}
	\end{align}

Similarly,
we denote the first time point that $\frac{\max\limits_{j=1,\cdots,n}\{\phi_j\}M(E_1(0))}{e^{\epsilon(T_1^I-\tau)}} = 1$ as $T_1^I$, and $\sup \limits_{t-\tau\le s\le t}\big(\max\limits_{j=1,\cdots,n}|e_j^I(s)|\big) \le 1$ for $t \ge T_1^I$, here
	\begin{align}
	T_1^I = {\epsilon}^{-1}\ln{\bigg(\max\limits_{j=1,\cdots,n}\{\phi_j\}M(E_1(0))\bigg)} + \tau.\label{t1i}
	\end{align}
	
Denote $T_1=\max(T_1^R, T_1^I)$, the absolute values of real-valued error systems (\ref{errorR}) and (\ref{errorI}) are all no more than than $1$ for $t \ge T_1$. It completes the first part of the proof.
	
{\bf{\it In the following, we prove the values of error systems will flow from $1$ to $0$ no more than time $T_2$.}}
	
	Pick two small constants $\rho^\ast, \rho^\star>0$ such that for all $j=1,\cdots,n$,
\begin{align}
	0<\rho^\ast<\xi_j^{-1}\bigg\{\widebar{\rho}_j-\bigg(&-d_{j}+(\{a_{jj}^R\}^{+}, \{-a_{jj}^I\}^{+})(\lambda_{j}^{RR},\lambda_{j}^{IR})^T\nonumber\\
&+(\phi_j^{-1}\xi_j)^{\frac{1}{\beta-1}}(|a_{jj}^R|, |a_{jj}^I|)(\lambda_j^{RI},\lambda_{j}^{II})^T\\
&+\xi_{j}^{\frac{1}{\beta-1}}\sum_{k\ne j}|\widehat a^T_{jk}|\widebar\Lambda_k(\xi_k^{\frac{1}{1-\beta}},\phi_k^{\frac{1}{1-\beta}})^T-\widebar \mu_{j}\bigg)^{+}\bigg\}\label{rhoast}
\end{align}
and
\begin{align}	0<\rho^\star<\phi_j^{-1}\bigg\{\widetilde{\rho}_j-\bigg(&-d_{j}+(\{a_{jj}^I\}^{+},\{a_{jj}^R\}^{+})(\lambda_{{j}}^{RI},\lambda_{{j}}^{II})^T\nonumber\\
&+(\xi_j^{-1}\phi_j)^{\frac{1}{\beta-1}}(|a_{jj}^I|,|a_{jj}^R|)(\lambda_{{j}}^{RR},\lambda_{{j}}^{IR})^T\\
&+\phi_{j}^{\frac{1}{\beta-1}}\sum_{k\ne j}|\widehat a^T_{jk}|\widetilde\Lambda_k(\xi_k^{\frac{1}{1-\beta}},\phi_k^{\frac{1}{1-\beta}})^T-\widetilde \mu_{j}\bigg)^{+}\bigg\}\label{rhostar}
\end{align}
	
Denote
	\begin{align}
	\rho = \min(\rho^\ast, \rho^\star).
	\end{align}
	
For all $t\ge T_1$, denote
\begin{align*}
E_2(t) =\bigg(\frac{e_1^R(t)^{1-\beta}}{1-\beta}, \cdots, \frac{e_n^R(t)^{1-\beta}}{1-\beta},\frac{e_1^I(t)^{1-\beta}}{1-\beta}, \cdots, \frac{e_n^I(t)^{1-\beta}}{1-\beta}\bigg)^T
	\end{align*}
with
	\begin{align}\label{acuteE}
	\|E_2(t)\|_{\{\xi,\infty\}} = \max\Big\{\max \limits_{j=1,\cdots,n}\big\{\xi_j^{-1}\frac{|e_j^R(t)|^{1-\beta}}{1-\beta}\big\},\max \limits_{j=1,\cdots,n}\big\{\phi_j^{-1}\frac{|e_j^I(t)|^{1-\beta}}{1-\beta}\big\}\Big\}
	\end{align}
	and
	\begin{align}\label{Vt}
	V(E_2(t)) = \sup \limits_{t-\tau\le s\le t}\big(\|E_2(s)\|_{\{\xi,\infty\}}+\rho s\big)
	\end{align}
Obviously, $\xi_j^{-1}\frac{|e_j^R(t)|^{1-\beta}}{1-\beta}+\rho t \le V(E_2(t))$ and $\phi_j^{-1}\frac{|e_j^I(t)|^{1-\beta}}{1-\beta}+\rho t \le V(E_2(t))$.
	
Similar to the procedure in the first part, we will first discuss the case that $\xi_j^{-1}\frac{|e_j^R(t)|^{1-\beta}}{1-\beta}+\rho t \le V(E_2(t)), j=1,2,\cdots,n$.
	
$(I)$ If $\xi_j^{-1}\frac{|e_j^R(t)|^{1-\beta}}{1-\beta}+\rho t < V(E_2(t))$, there must be a constant $\delta_2>0$ such that $\xi_j^{-1}\frac{|e_j^R(s)|^{1-\beta}}{1-\beta}+\rho s < V(E_2(t))$, and $V(E_2(s)) \le V(E_2(t))$ for $s \in (t, t+\delta_2)$.
	
$(II)$ If there exist an index $\widebar j_1$ and a time point $\widebar t_1\ge T_1$ such that $\xi_{\widebar j_1}^{-1}\frac{|e_{\widebar j_1}^R({\widebar t_1})|^{1-\beta}}{1-\beta}+\rho {\widebar t_1} = V(E_2({\widebar t_1}))$, then we have
\begin{align}
&\xi_{\widebar j_1}\frac{dV(E_2(t))}{dt}\bigg|_{t=\widebar t_1}=\frac{d}{dt}\bigg(\frac{|e_{\widebar j_1}^R(t)|^{1-\beta}}{1-\beta}+\xi_{\widebar j_1}\rho t\bigg)\bigg|_{t=\widebar t_1}\nonumber\\
	\le
	&|e_{\widebar j_1}^R(\widebar t_1)|^{-\beta}\bigg\{-d_{\widebar j_1}|e_{\widebar j_1}^R(\widebar t_1)|+(\{a_{{\widebar j_1}{\widebar j_1}}^R\}^{+}, \{-a_{{\widebar j_1}{\widebar j_1}}^I\}^{+})(\lambda_{{\widebar j_1}}^{RR},\lambda_{{\widebar j_1}}^{IR})^T|e_{\widebar j_1}^R(\widebar t_1)|\nonumber\\
&+(|a_{{\widebar j_1}{\widebar j_1}}^R|, |a_{{\widebar j_1}{\widebar j_1}}^I|)(\lambda_{{\widebar j_1}}^{RI},\lambda_{{\widebar j_1}}^{II})^T|e_{\widebar j_1}^I(\widebar t_1)|+\sum_{k\ne {\widebar j_1}}|\widehat a^T_{{\widebar j_1}k}|\widebar\Lambda_k|\widehat e_k(\widebar t_1)|\nonumber\\	
&+\sum_{k=1}^n|\widehat b^T_{{\widebar j_1}k}|\widebar \Gamma_k |\widehat e_{k\tau_{\widebar j_1k}}(\widebar t_1)|+2|H_{\widebar j_1}^R| -\widebar \mu_{\widebar j_1}|e_{\widebar j_1}^R(\widebar t_1)| - \widebar \rho_{\widebar j_1}{|e_{\widebar j_1}^R(\widebar t_1)|}^\beta-\widebar\eta_{\widebar j_1}^R\bigg\}\nonumber\\
&+\xi_{\widebar j_1}\rho\label{add9}
	\end{align}
	
From (\ref{acuteE}), we have
	\begin{align*}
	\xi_k^{-1}\frac{|e_k^R(\widebar t_1)|^{1-\beta}}{1-\beta} \le \xi_{\widebar j_1}^{-1}\frac{|e_{\widebar j_1}^R(\widebar t_1)|^{1-\beta}}{1-\beta},
	\end{align*}
	i.e.,
\begin{align}\label{add6}
	\xi_k^{\frac{1}{\beta-1}}|e_k^R(\widebar t_1)| \le \xi_{\widebar j_1}^{\frac{1}{\beta - 1}}|e_{\widebar j_1}^R(\widebar t_1)|, ~~~k=1,\cdots,n
\end{align}

Moreover, note that $\sup \limits_{T_1-\tau \le s\le T_1}\big(\max\limits_{j=1,\cdots,n}|e_j^R(s)|\big) \le 1$, and as long as\\ $\sup \limits_{t-\tau \le s \le t}\big(\max\limits_{j=1,\cdots,n}|e_j^R(s)|\big) \le 1$, then
\begin{align}\label{add7}
|e_j^R(t-\tau_{jk}(t))|\le 1,~~~j,k=1,\cdots,n
\end{align}
	
Similarly, we can also get that
\begin{align}\label{add8}
\phi_k^{\frac{1}{\beta-1}}|e_k^I(\widebar t_1)| \le \xi_{\widebar j_1}^{\frac{1}{\beta - 1}}|e_{\widebar j_1}^R(\widebar t_1)|.
\end{align}

Therefore, combining with (\ref{add9})-(\ref{add8}), one can get
	\begin{align*}
	&\xi_{\widebar j_1}\frac{dV(E_2(t))}{dt}\bigg|_{t=\widebar t_1}\\
	\le
	&|e_{\widebar j_1}^R(\widebar t_1)|^{-\beta}\bigg\{-d_{\widebar j_1}|e_{\widebar j_1}^R(\widebar t_1)|+(\{a_{{\widebar j_1}{\widebar j_1}}^R\}^{+}, \{-a_{{\widebar j_1}{\widebar j_1}}^I\}^{+})(\lambda_{{\widebar j_1}}^{RR},\lambda_{{\widebar j_1}}^{IR})^T|e_{\widebar j_1}^R(\widebar t_1)|\nonumber\\
&+(|a_{{\widebar j_1}{\widebar j_1}}^R|, |a_{{\widebar j_1}{\widebar j_1}}^I|)(\lambda_{{\widebar j_1}}^{RI},\lambda_{{\widebar j_1}}^{II})^T(\phi_{\widebar j_1}^{-1}\xi_{\widebar j_1})^{\frac{1}{\beta-1}}|e_{\widebar j_1}^R(\widebar t_1)|\\
&+\sum_{k\ne {\widebar j_1}}|\widehat a^T_{{\widebar j_1}k}|\widebar\Lambda_k\mathrm{diag}(\xi_k^{\frac{1}{1-\beta}},\phi_k^{\frac{1}{1-\beta}})\bigg(\xi_k^{\frac{1}{\beta-1}}|e_k^R(\widebar t_1)|,\phi_k^{\frac{1}{\beta-1}}|e_k^I(\widebar t_1)|\bigg)^T\nonumber\\	
&+\sum_{k=1}^n|\widehat b^T_{{\widebar j_1}k}|\widebar \Gamma_k (1,1)^T +2|H_{\widebar j_1}^R| -\widebar \mu_{\widebar j_1}|e_{\widebar j_1}^R(\widebar t_1)| - \widebar \rho_{\widebar j_1}{|e_{\widebar j_1}^R(\widebar t_1)|}^\beta-\widebar\eta_{\widebar j_1}^R\bigg\}\nonumber\\
&+\xi_{\widebar j_1}\rho\\
\le
	&|e_{\widebar j_1}^R(\widebar t_1)|^{-\beta}\bigg\{\bigg(-d_{\widebar j_1}+(\{a_{{\widebar j_1}{\widebar j_1}}^R\}^{+}, \{-a_{{\widebar j_1}{\widebar j_1}}^I\}^{+})(\lambda_{{\widebar j_1}}^{RR},\lambda_{{\widebar j_1}}^{IR})^T\nonumber\\
&+(\phi_{\widebar j_1}^{-1}\xi_{\widebar j_1})^{\frac{1}{\beta-1}}(|a_{{\widebar j_1}{\widebar j_1}}^R|, |a_{{\widebar j_1}{\widebar j_1}}^I|)(\lambda_{{\widebar j_1}}^{RI},\lambda_{{\widebar j_1}}^{II})^T\\
&+\xi_{\widebar j_1}^{\frac{1}{\beta-1}}\sum_{k\ne {\widebar j_1}}|\widehat a^T_{{\widebar j_1}k}|\widebar\Lambda_k(\xi_k^{\frac{1}{1-\beta}},\phi_k^{\frac{1}{1-\beta}})^T-\widebar \mu_{\widebar j_1}\bigg)|e_{\widebar j_1}^R(\widebar t_1)|\nonumber\\	
&+\bigg(\sum_{k=1}^n|\widehat b^T_{{\widebar j_1}k}|\widebar \Gamma_k (1,1)^T +2|H_{\widebar j_1}^R|-\widebar\eta_{\widebar j_1}^R\bigg)\bigg\}+\bigg(- \widebar \rho_{\widebar j_1}+\xi_{\widebar j_1}\rho\bigg)\\
\le
	&\bigg(-d_{\widebar j_1}+(\{a_{{\widebar j_1}{\widebar j_1}}^R\}^{+}, \{-a_{{\widebar j_1}{\widebar j_1}}^I\}^{+})(\lambda_{{\widebar j_1}}^{RR},\lambda_{{\widebar j_1}}^{IR})^T\nonumber\\
&+(\phi_{\widebar j_1}^{-1}\xi_{\widebar j_1})^{\frac{1}{\beta-1}}(|a_{{\widebar j_1}{\widebar j_1}}^R|, |a_{{\widebar j_1}{\widebar j_1}}^I|)(\lambda_{{\widebar j_1}}^{RI},\lambda_{{\widebar j_1}}^{II})^T\\
&+\xi_{\widebar j_1}^{\frac{1}{\beta-1}}\sum_{k\ne {\widebar j_1}}|\widehat a^T_{{\widebar j_1}k}|\widebar\Lambda_k(\xi_k^{\frac{1}{1-\beta}},\phi_k^{\frac{1}{1-\beta}})^T-\widebar \mu_{\widebar j_1}\bigg)^{+}+\bigg(- \widebar \rho_{\widebar j_1}+\xi_{\widebar j_1}\rho\bigg)<0
	\end{align*}
which implies that there must exist $\sigma_2>0$ such that $\xi_j^{-1}\frac{|e_j^R(s)|^{1-\beta}}{1-\beta}+\rho s <\xi_{\widebar j_1}^{-1}\frac{|e_{\widebar j_1}^R({\widebar t_1})|^{1-\beta}}{1-\beta}+\rho {\widebar t_1}$ holds for all $s\in ({\widebar t_1},{\widebar t_1}+\sigma_2)$.
	
	For the other condition $\phi_j^{-1}\frac{|e_j^I(t)|^{1-\beta}}{1-\beta}+\rho t \le V(E_2(t))$, it can be analysed in the same way, so it is omitted here.
	
	Thus, we conclude that
	\begin{align*}
	\min\limits_{j=1,\cdots,n}\{\xi_j^{-1}\}\max\limits_{j=1,\cdots,n}\frac{|e_j^R(t)|^{1-\beta}}{1-\beta}+\rho t \le V(E_2(t)) \le V(E_2(T_1))
	\end{align*}
	\begin{align*}
	\min\limits_{j=1,\cdots,n}\{\phi_j^{-1}\}\max\limits_{j=1,\cdots,n}\frac{|e_j^I(t)|^{1-\beta}}{1-\beta}+\rho t \le V(E_2(t)) \le V(E_2(T_1))
	\end{align*}
	i.e.,
	\begin{align*}
	\max\limits_{j=1,\cdots,n}|e_j^R(t)|^{1-\beta}&\le& (1-\beta)\max\limits_{j=1,\cdots,n}\xi_j\cdot\Big(\sup\limits_{T_1-\tau\le s\le T_1}\|E_2(s)\|_{\{\xi,\infty\}}-\rho (t-T_1)\Big)\\
	\max\limits_{j=1,\cdots,n}|e_j^I(t)|^{1-\beta}&\le& (1-\beta)\max\limits_{j=1,\cdots,n}\phi_j\cdot\Big(\sup\limits_{T_1-\tau\le s\le T_1}\|E_2(s)\|_{\{\xi,\infty\}}-\rho (t-T_1)\Big)
	\end{align*}
It is obvious that $\max\limits_{j=1,\cdots,n}|e_j^R(t)|$ and $\max\limits_{j=1,\cdots,n}|e_j^I(t)|$ will decrease to $0$, denote $T_2$ as the first time they all become $0$, then $\max\limits_{j=1,\cdots,n}|e_j^R(T_2)|^{1-\beta}=0$ and $\max\limits_{j=1,\cdots,n}|e_j^I(T_2)|^{1-\beta}=0$. Thus we obtain
	\begin{align}
	T_2= \frac{1}{\min\{\xi\}\cdot\rho(1-\beta)} + T_1
	\end{align}
	here $\min\{\xi\}$ = $\min\{\min\limits_{j=1,\cdots,n}\{\xi_j\}, \min\limits_{j=1,\cdots,n}\{\phi_j\}\}$, which means that, the absolute value of real-valued error systems will flow from $1$ to $0$ no longer than $T_2$. It completes the proof.
\endproof

\begin{rem}
If we do not consider the effect of sign, i.e., the condition for $f$ in (\ref{add3}) is replaced by the Lipschitz condition, i.e.,
\begin{align}
\bigg|\frac{\partial{f_k^{\ell_1}}(x_k^R,x_k^I)}{\partial{x_k^{\ell_2}}}\bigg|\le \lambda_k^{\ell_1\ell_2},
\end{align}
then the conditions (\ref{mubar})-(\ref{rhotilde}) in Theorem 1 are replaced by the following:
\begin{align}
	\widebar{\mu}_j &>-d_j+\xi_j^{-1}\Big(\sum_{k=1}^{n}|\widehat a_{jk}^T|\widebar\Lambda_k (\xi_k,\phi_k)^T+\sum_{k=1}^n|\widehat b_{jk}^T|\widebar \Gamma_k (\xi_k,\phi_k)^T\Big)\label{mubar1}\\
	\widetilde{\mu}_j &>-d_j+\phi_j^{-1}\Big(\sum_{k=1}^{n}|\widehat a_{jk}^T|\widetilde\Lambda_k (\xi_k,\phi_k)^T+\sum_{k=1}^n|\widehat b_{jk}^T|\widetilde \Gamma_k (\xi_k,\phi_k)^T\Big)\label{mutilde1}\\
	\widebar{\rho}_j&>max\Big(0,-d_j+\xi_j^{\frac{1}{\beta-1}}\sum_{k=1}^n|\widehat a_{jk}^T|\widebar\Lambda_k(\xi_k^{\frac{1}{1-\beta}},\phi_k^{\frac{1}{1-\beta}})^T-\widebar \mu_j\Big)\label{rhobar1}\\
	\widetilde{\rho}_j&>max\Big(0,-d_j+\phi_j^{\frac{1}{\beta-1}}\sum_{k=1}^n|\widehat a_{jk}^T|\widetilde \Lambda_k (\xi_k^{\frac{1}{1-\beta}},\phi_k^{\frac{1}{1-\beta}})^T-\widetilde \mu_j\Big)\label{rhotilde1}
	\end{align}	
\end{rem}

\begin{rem}
When $\xi=(1,\cdots,1)^T$ and the CVNN is RVNN, then it becomes the case discussed in \cite{wang18}, so this paper can be regarded as a generalization of the result in \cite{wang18}.
\end{rem}

\begin{rem}
In fact, the problem can be solved without using the matrix representation, but the advantage of the matrix method is that it can be easier to be extended to higher dimension neural networks, such as quaternion-valued neural networks \cite{liuli2019}.
\end{rem}

\begin{rem}
SYN can also be solved by the same process as in this theorem, and in some aspect, the process is easier than A-SYN, interested readers are encouraged to complete the proof.
\end{rem}

\begin{rem}
In this paper, we deal with the A-SYN by decomposing the CVNN into RVNNs, in order to compensate the condition that the time-varying delay is asynchronous. In fact, we can also solve this paper by regarding the CVNN as a whole, but in this case, the time-varying delay should be restrict to be the same, which will be considered in our following papers.
\end{rem}

\section{Numerical simulations}\label{nu}
	In this part, a numerical example is given to show the correctness of our results.

	Consider a two-neuron master-slave CVNN described as follows:
	\begin{align}\label{equ1}
	\left\{
	\begin{array}{ll}
	\dot{x}_1(t)=&-d_1x_1(t)+a_{11}f_1(x_1(t))+a_{12}f_2(x_2(t))\\
	&+b_{11}g_1(x_1(t-\tau_{11}(t)))+b_{12}g_2(x_2(t-\tau_{12}(t)))+H_1\\
	\dot{x}_2(t)=&-d_2x_2(t)+a_{21}f_1(x_1(t))+a_{22}f_2(x_2(t))\\
	&+b_{21}g_1(x_1(t-\tau_{21}(t)))+b_{22}g_2(x_2(t-\tau_{22}(t)))+H_2\\
	\dot{y}_1(t)=&-d_1y_1(t)+a_{11}f_1(y_1(t))+a_{12}f_2(y_2(t))\\
	&+b_{11}g_1(y_1(t-\tau_{11}(t)))+b_{12}g_2(y_2(t-\tau_{12}(t)))+H_1+u_1\\
	\dot{y}_2(t)=&-d_2y_2(t)+a_{21}f_1(y_1(t))+a_{22}f_2(y_2(t))\\
	&+b_{21}g_1(y_1(t-\tau_{21}(t)))+b_{22}g_2(y_2(t-\tau_{22}(t)))+H_2+u_2
	\end{array}
	\right.
	\end{align}
	where $x_j=x_j^R+ix_j^I$, $y_j=y_j^R+iy_j^I$, $j=1,2$, $d_1=0.5$, $d_2=1$,
	\begin{align*}
	&A=(a_{jk})_{2\times 2}=\left(\begin{array}{cc}1.2+0.2i&0.8+1.2i\\1+1.5i&0.4+0.2i\end{array}\right),\\
	&B=(b_{jk})_{2\times 2}=\left(\begin{array}{cc}0.2+1.2i&0.2+0.8i\\1.5+i&0.2+0.4i\end{array}\right),\\
&f_k(x_k)=\frac{1-\mathrm{exp}(-x_k^R-2x_k^I)}{1+\mathrm{exp}(-x_k^R-2x_k^I)}+i\frac{1-\mathrm{exp}(-2x_k^R-x_k^I)}{1+\mathrm{exp}(-2x_k^R-x_k^I)},\\
	&g_k(x_k)=\frac{|x_k^R+x_k^I+1|-|x_k^R+x_k^I-1|}{2}+i\frac{|x_k^R+x_k^I+1|-|x_k^R+x_k^I-1|}{2},\\
	&\tau_{11}=\frac{e^t}{1+e^t},~~\tau_{12}=\frac{e^t-0.5}{1+e^t},~~\tau_{21}=\frac{1}{1+|\cos(10t)|}, ~~\tau_{22}=\frac{1}{1+|\sin(10t)|},\\
	&\mathrm{obviously}~\tau_{jk}(t) \le \tau=1~\mathrm{for}~j,k = 1,\cdots, n,\\
	&H_1=0.1+0.1i,\\
&H_2=0.2+0.2i,\\
	&\Phi(\theta)=(\Phi_1(\theta),\Phi_2(\theta))^T=(-1-2i,1.5-1.5i)^T, \theta\in [-1,0],\\
	&\Psi(\theta)=(\Psi_1(\theta),\Psi_2(\theta))^T=(5+5.4i,-5.4-3.5i)^T, \theta\in [-1,0]
	\end{align*}
	and with some simple calculations, we have
	\begin{align*}
	\widebar\Lambda_k = \left(\begin{array}{cc}
	0.5&1\\1&0.5
	\end{array}\right),
	\widetilde\Lambda_k = \left(\begin{array}{cc}
	1&0.5\\0.5&1
	\end{array}\right),
	\widebar\Gamma_k=\widetilde \Gamma_k = \left(\begin{array}{cc}
	1&1\\1&1
	\end{array}\right),~~~k = 1,2
	\end{align*}

	\begin{figure}
		\centering
		\includegraphics[height=0.4\textheight,width=0.8\textwidth]{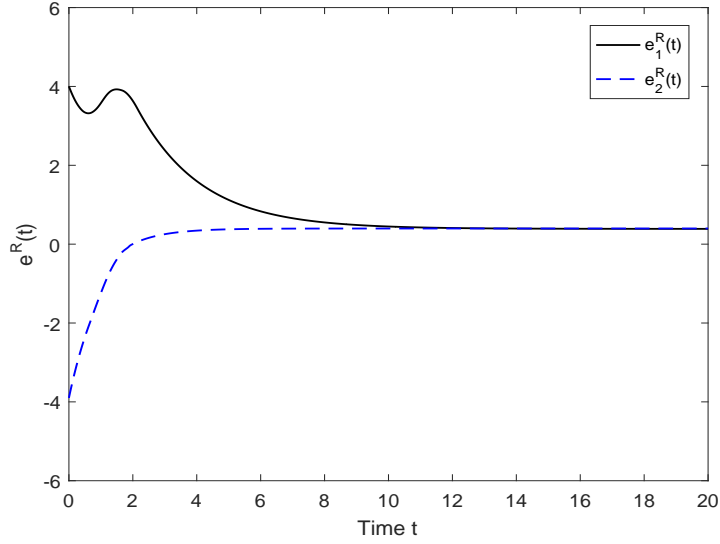}
		\caption{Real part trajectories of error system (\ref{equ1}) without control.}
		\label{fig1}
	\end{figure}
	\begin{figure}
		\centering
		\includegraphics[height=0.4\textheight,width=0.8\textwidth]{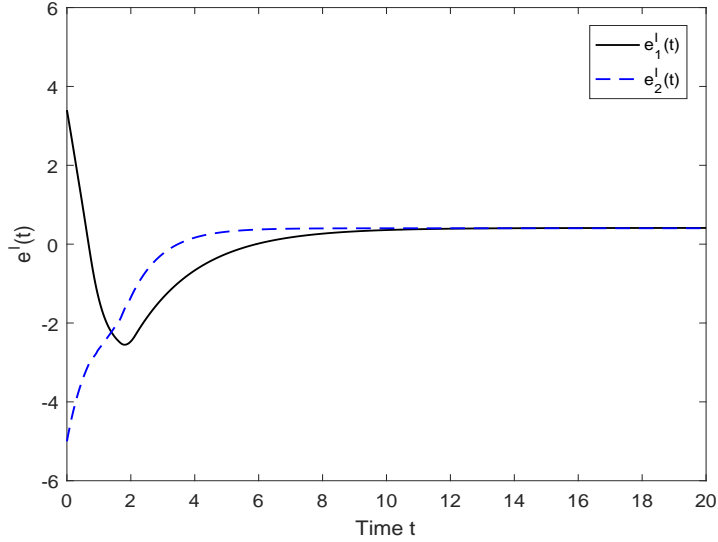}
		\caption{Imaginary part trajectories of error system (\ref{equ1}) without control.}
		\label{fig2}
	\end{figure}
	
Figures \ref{fig1} and \ref{fig2} show the trajectories of error system (\ref{equ1}) without control, as time increases, it is obvious that system cannot achieve anti-synchronization even they are at the equilibrium point.
	
Then we choose $\xi = (\xi_1,\xi_2,\phi_1,\phi_2)^T=(0.4,0.8,0.5,0.6)^T$, $\beta = 0.5$, and from calculations, inequalities (\ref{mubar})-(\ref{etatilde}) are:
\begin{align*}
&\widebar\mu_{1}>14.675, \widetilde \mu_{1}>11.9,&&\widebar\mu_{2}>7.531,\widetilde\mu_{2}>10.008,\\
&\widebar\rho_{1}>(12.681-\widebar\mu_{1})^+=0, &&\widetilde\rho_{1}>(8.244-\widetilde \mu_{1})^+=0,\\
&\widebar \rho_{2}>(2.665-\widebar\mu_{2})^+=0, &&\widetilde\rho_{2}>(4.393-\widetilde \mu_{2})^+=0,\\
&\widebar\eta_{1}\ge 5,~~~~\widetilde\eta_{1}\ge 5, &&\widebar \eta_{2}\ge 6.6, ~~~~\widetilde\eta_{2}\ge 6.6
\end{align*}
as a result, the control scheme can be defined as follows,
	\begin{align}\label{control1}
	\left\{
	\begin{array}{ll}
	u_1^R=&-\mathrm{sign}(e_1^R(t))\big[18|e_1^R(t)|+0.2|e_1^R(t)|^{0.5}+5\big],\\
	u_1^I=&-\mathrm{sign}(e_1^I(t))\big[15|e_1^I(t)|+0.2|e_1^I(t)|^{0.5}+5\big],\\
	u_2^R=&-\mathrm{sign}(e_2^R(t))\big[10|e_2^R(t)|+0.4|e_2^R(t)|^{0.5}+6.6\big],\\
	u_2^I=&-\mathrm{sign}(e_2^I(t))\big[12|e_2^I(t)|+0.4|e_2^I(t)|^{0.5}+6.6\big]
	\end{array}
	\right.
	\end{align}
	
\begin{figure}
		\centering
		\includegraphics[height=0.4\textheight,width=0.8\textwidth]{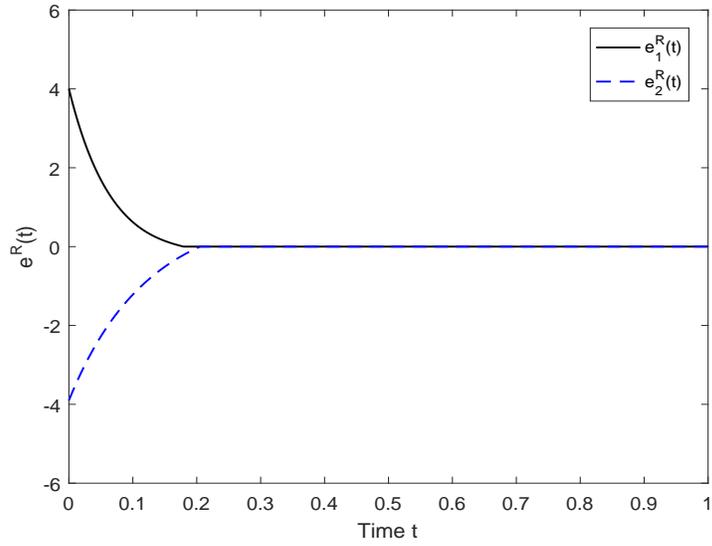}
		\caption{Real part trajectories of error system (\ref{equ1}) under control scheme (\ref{control1}).}
		\label{fig3}
	\end{figure}

\begin{figure}
		\centering
		\includegraphics[height=0.4\textheight,width=0.8\textwidth]{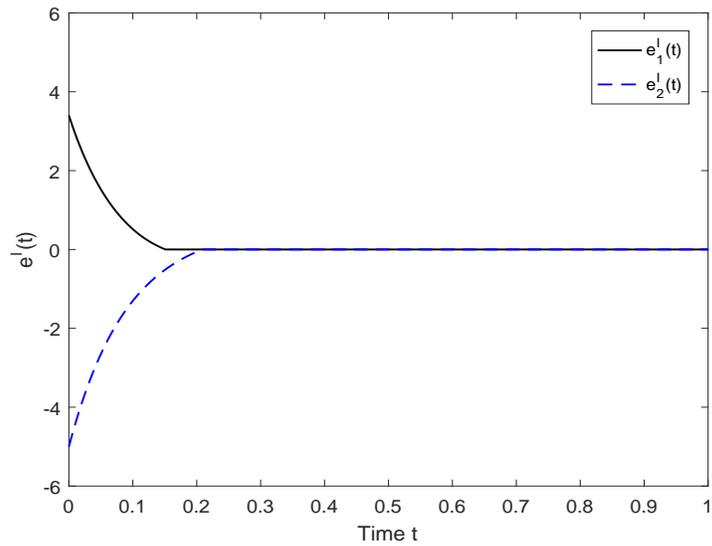}
		\caption{Imaginary part trajectories of error system (\ref{equ1}) under control scheme (\ref{control1}).}
		\label{fig4}
	\end{figure}

\begin{figure}
		\centering
		\includegraphics[height=0.43\textheight,width=0.8\textwidth]{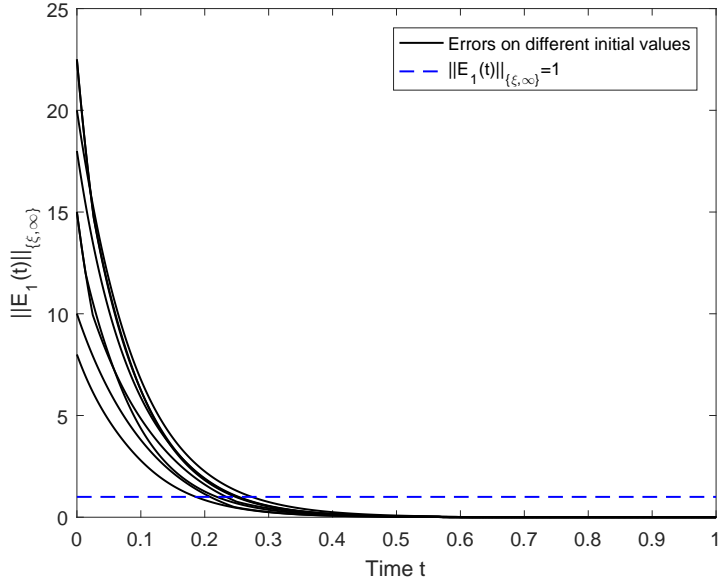}
		\caption{Imaginary part trajectories of error system (\ref{equ1}) under control scheme (\ref{control1}).}
		\label{fig5}
	\end{figure}

\begin{figure}
		\centering
		\includegraphics[height=0.4\textheight,width=0.8\textwidth]{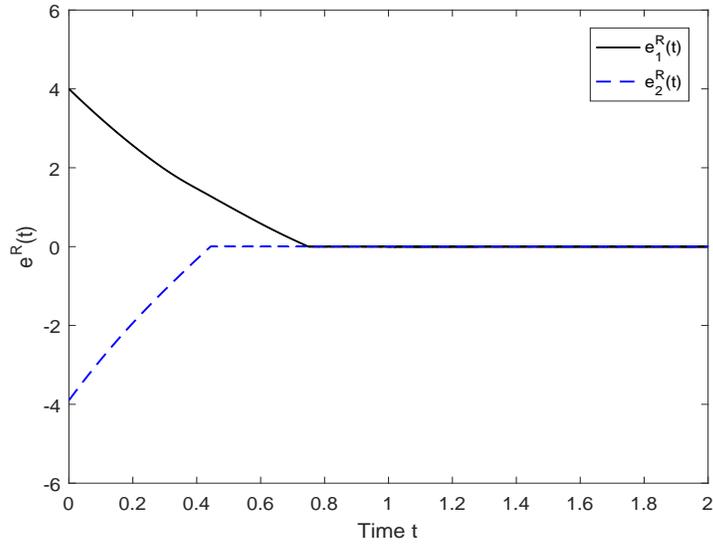}
		\caption{Real part trajectories of error system (\ref{equ1}) under weaker control parameter.}
		\label{fig6}
	\end{figure}

\begin{figure}
		\centering
		\includegraphics[height=0.4\textheight,width=0.8\textwidth]{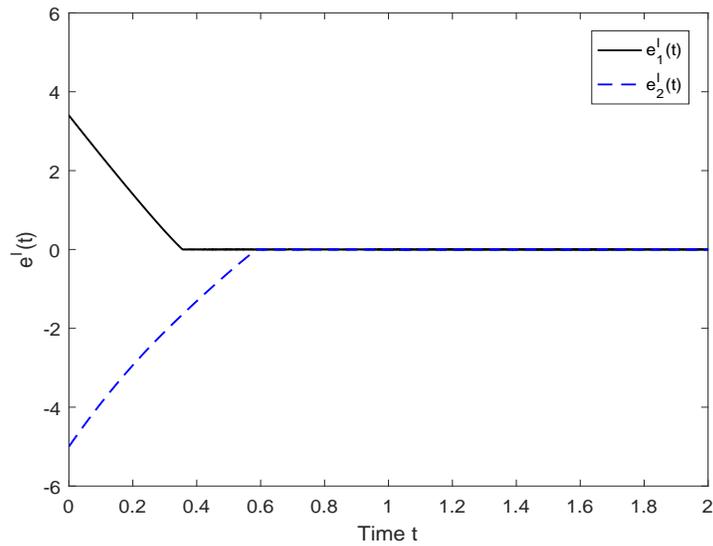}
		\caption{Imaginary part trajectories of error system (\ref{equ1}) under weaker control parameter.}
		\label{fig7}
	\end{figure}

	Figures \ref{fig3} and \ref{fig4} show trajectories of error system (\ref{equ1}) with above control, we can see that error system reaches anti-synchronization in finite time.
	In our proof part, we use the $\{\xi, \infty\}$-norm as a measure, the defined error function (\ref{xtinfty}) will flow from initial value to $1$ in finite time, then decrease to $0$ in fixed time theoretically. Figure \ref{fig5} shows the trajectories of $\{\xi, \infty\}$-norm errors under different random initial values.
	Actually, as we have discussed in the proof process, the theoretical finite time $T_1$ and $T_2$ can be calculated directly. Pick $\epsilon = 0.25$, $\rho=0.4$, which makes inequalities (\ref{first1}), (\ref{rhoast}), and (\ref{rhostar}) holds, from (\ref{t1r}) and (\ref{t1i}), we have
	\begin{align*}
&T_1 = T_1^R = 0.25^{-1}\ln(0.8*10)+1 = 9.318,	\\
&T_2 =\frac{1}{0.4\cdot 0.4\cdot (1-0.5)} + T_1=21.818
	\end{align*}
which means that system (\ref{equ1}) will achieve finite time A-SYN no longer than $T_2$. However, we find that there is some distance between the theoretical result and the practical one.
That is to say, the intensity of control (\ref{control1}) is too high and can be smaller while system (\ref{equ1}) can still achieve finite time A-SYN. Figures \ref{fig6} and \ref{fig7} show the trajectories under the following control with parameters $\widebar\mu_{1}=0.18, \widetilde \mu_{1}=0.15, \widebar\rho_{1}=\widetilde\rho_{1}=0.02, \widebar\mu_{2}=0.1, \widetilde \mu_{2}=0.12, \widebar\rho_{1}=\widetilde\rho_{1}=0.04$ and other parameters are not changed.
\section{Conclusion}\label{conclude}
In this paper, the A-SYN problem for CVNNs with bounded and asynchronous time delays is investigated. By decomposing the CVNNs toto multi equivalent RVNNs and utilizing the $\{\xi,\infty\}$-norm, we give some sufficient criteria for A-SYN. Finally, in numerical simulation part, we present an example to show the validity of these obtained criteria. It is worth noting that, the method using in this paper can be extended to solve both SYN and A-SYN problem for higher dimensional neural networks, such as quaternion-valued neural networks, the generalized norm also has some other choices, and we can also solve the problem by regarding the CVNN as a whole rather than decomposing it into several parts. These will be studied in our future work.

\end{document}